\documentclass[preprint,showpacs,prl,superscriptaddress]{revtex4}
\usepackage{graphicx}  
\usepackage{natbib}    

\begin{document}

\title{Experimental investigation of laminar turbulent intermittency in pipe flow}

\author{Devranjan Samanta, Alberto de Lozar and Bj\"orn Hof}
\affiliation{Max Plank Institute for Dynamics and Self-Organisation, G\"ottingen, 37073 ,Germany}

\begin{abstract}
In shear flows turbulence first occurs in the form of localized structures (puffs/spots) surrounded by laminar fluid. We here investigate such spatially intermittent flows in a pipe experiment showing that turbulent puffs have a well defined interaction distance, which sets their minimum spacing as well as the maximum observable turbulent fraction. Two methodologies are employed. Starting from a laminar flow, puffs are first created by locally injecting a jet of fluid through the pipe wall. When the perturbation is applied periodically at low frequencies, as expected, a regular sequence of puffs is observed where the puff spacing is given by the ratio of the mean flow speed to the perturbation frequency. At large frequencies however puffs are found to interact and annihilate each other. Varying the perturbation frequency an interaction distance is determined which sets the highest possible turbulence fraction. This enables us to establish an upper bound for the friction factor in the transitional regime which provides a well defined link between the Blasius and the Hagen-Poiseuille friction laws. In the  second set of experiments, the Reynolds number is reduced suddenly from fully turbulent to the intermittent regime.The resulting flow reorganizes itself to a sequence of constant size puffs which, unlike in Couette and Taylor Couette flow are randomly spaced. The minimum distance between the turbulent patches is identical to the puff interaction length. The puff interaction length is found to be in agreement with the wavelength of regular stripe and spiral patterns in plane Couette and Taylor-Couette flow. 

\end{abstract}

\maketitle

\section{Introduction}
Recently substantial progress has been made in understanding the dynamics underlying turbulence in pipe flow close to onset \cite[]{Kerswell,Eckhardt2}.  A large number of unstable solutions to the problem have been identified (mostly travelling waves) and it has been suggested that the turbulent dynamics evolve around such unstable states. Although similar coherent states have been observed in experiments \cite[]{Hof4}, an important qualitative difference between the exact solutions observed to date and structures within the turbulent flow is that while the former ones are spatially periodic the latter are localized in the streamwise direction.   Specifically at low Reynolds numbers (Re$\approx$2000) turbulence only occurs in localized patches \cite[]{Reynolds, Rotta} also referred to as equilibrium puffs \cite[]{Wygnanski1, Wygnanski2},  which are separated by laminar fluid. Only when the Reynolds number is further increased turbulent patches begin to spatially expand. A better understanding of the processes underlying localisation \cite[]{Schneider} is an essential requirement to allow further theoretical progress. This regime is also of practical interest since the pressure and flow rate fluctuations in the laminar-turbulent intermitent flows are very large (markedly higher than in the case of a fully turbulent flow).  In addition edge states \cite[]{Mellibovsky, Willis, Duguet} separating laminar from turbulent flow and which are potentially relevant for the transition process remain bounded in axial direction ever at much higher Re. This indicates that even beyond the intermittent regime localisation is relevant to some aspects of the flow.

Localized turbulence can be identified in a number of canonical shear flows, but unlike in pipe and channel flows, for Taylor-Couette (TCF) and plane Couette flow (PCF) regular spatio temporal structures have been reported to consist of tilted laminar-turbulent stripes. In the case of Taylor-Couette flow the localized turbulent stripes are traditionally known as turbulent spirals \cite[]{Coles}. The existence of a distinct preferred wavelength has led to the speculation of a wavelength instability preceding turbulence \cite[]{Prigent, Prigent1, Barkley1, Cros}. In pipes, on the other hand, many earlier investigations of the intermittent regime \cite[]{Rotta, Wygnanski1} were  focused on quantities like intermittency factors and turbulence fraction which for observation times typically realized in experiments and simulations ($\sim$  a few hundred $D/U$), are strongly dependent on initial conditions  as well as imperfections, such as disturbances created at the pipe entrance. In this paper we quantify the spatial interaction in the intermittent regime and we aim to determine an upper bound on turbulence fractions and friction factors, which unlike above quantities do not depend on the specifics of the experimental setup. The resulting puff interaction distance is then compared to puff spacings which naturally occur after a sudden Reynolds number reduction from a fully turbulent flow. Finally the resulting puff spacing and interaction distances are compared to typical wavelength in intermittent Couette and Taylor Couette flows.

\begin{figure}
\centering
\includegraphics[width=13.5cm]{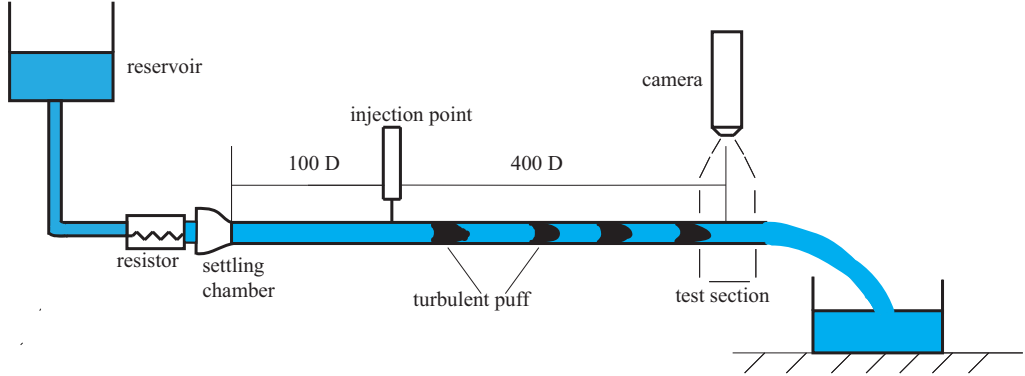}
\vspace{-1.5em}
\caption{Schematic of the experimental setup}
\vspace{-1.5em} 
\label{setup}
\end{figure}

\section{Experimental setup} 
The experimental set up consists of a long glass pipe of D=10 $\pm$ 0.01 mm diameter. The pipe is made from 1 meter long sections joined to form the total length of 6 meters. The junctions are carefully machined to avoid disturbances in the flow. The entrance consists of a 200 mm long cylinder with a diameter of $D=100$ mm and contains 2 meshes at  the upstream end and a smooth convergence to the 10 mm pipe at the downstream end. In the absence of external perturbations flow remains laminar up to $Re= 8000$. Distilled water is used as the working fluid and formation of algae is avoided by adding 300 ppm of Sodium Hypochlorite.

The flow velocity is regulated by altering the height of a reservoir tank connected to the pipe. A pump is employed at the pipe end to recirculate the water to the reservoir and ensures that the pressure head is held constant. A sketch of the experimental setup is shown in figure \ref{setup}. The mean flow velocity can be determined by weighing the mass of discharged fluid over a period of time (typically 60 seconds). In order to reduce temperature fluctuations that alter the liquid viscosity the water is circulated through a heat exchanger before entering the pipe. The temperature of the water is continuously monitored by a PT 100 thermal probe which shows typical variations of 0.1 K over 3-4 hours. Repeated measurements show that Reynolds number variations are less than $\pm$ 0.5 $\%$ over a 3-4 hour period. When the flow changes from laminar to turbulent, the drag increases and consequently the flow rate in a gravity driven pipe is reduced. This undesired effect is minimized by circulating the water through a big flow resistance (in our case a tube of small diameter) before entering the pipe. Since $\sim$ 96 $\%$ of the pressure drop occurs over the entrance resistance, the mean flow is virtually unaltered by the increase of drag during transition in the pipe. Detailed flowrate measurements shows that in the present setup the Reynolds number decreases by less than 10 when the flow is changed from laminar to intermittent. If the flow is in a steady state (i.e. laminar or intermittent at a fixed turbulent fraction), so either before or shortly after such a reduction, instantaneous pressure measurments (Validyne DP45 low pressure transducer) show that fluctuations in the flowrate and hence in Re are even much smaller than $\Delta Re \pm 1$. The largest variations in Re are casued by long term temperature drifts (typically take place on order of hours) where a change by 0.1 K results in a Reynolds number change of about $\Delta Re=6$.

\begin{figure}
\centering
\includegraphics[width=13.5cm]{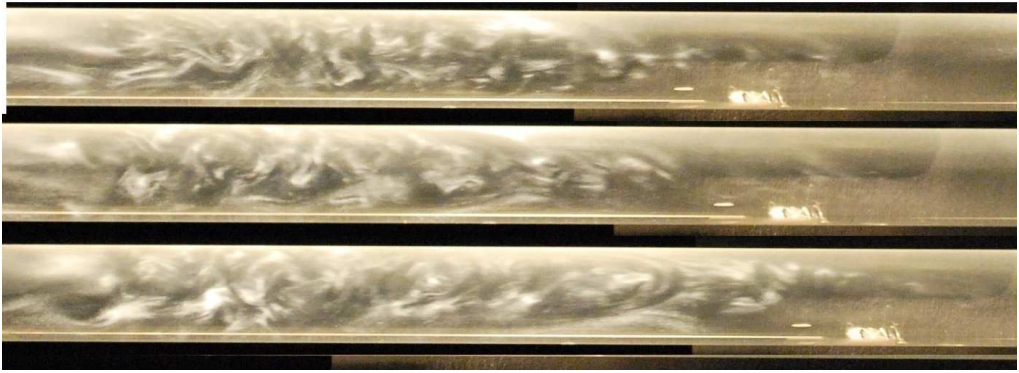}
\vspace{-1.5em}
\caption{ Flow visualisation images of a turbulent puff at Re = 2000 at 50 D (top), 100 D (middle) and 150 D (bottom) from the point of perturbation. Flow direction is from left to right. The puff reaches its typical size in less than 50 D from the perurbation point and the size is constant except for statistical fluctuations.}
\vspace{-0.5em} 
\label{puffpicture}
\end{figure}

 Five hundred D downstream of the pipe entrance, a segment of the pipe is illuminated by a thin light sheet (parallel to the pipe axis). Neutrally buoyant anisotropic particles (Mearlmaid AA) are used for flow visualization purposes. These particles have the form of flat platelets which align with the shear \cite[]{Savas, Gauthier} and hence turbulent flow can be easily distinguished from laminar motion. Turbulence can be triggered by a brief injection of a jet of water through a 1 mm hole in the pipe wall. As shown in figure \ref{puffpicture} at Reynolds number Re $\approx$ 2000 such a disturbance results in a localized turbulent puff which develops its typical shape in less than 50 advective time units. Images were taken with a 10 megapixel still camera and capture a pipe segment of about 10 D long. Throughout the paper non dimensional units are used for time and length scales unless stated otherwise. Space is given in pipe diameters (D), velocity in the mean flow velocities (U) and time in the corresponding unit ($D$/$U$). 

\section{Results} 
\subsection{Periodic perturbation} 

 The first set of experiments are concerned with determining the distance at which puffs first interact.  When puffs are located too close to each other the downstream puff decays \cite[]{Hof1}. To find the exact distance where this interaction starts periodic perturbations are applied to the flow at a fixed position and the rate of puff triggering as response of the injected disturbance is observed at a downstream distance. 

The perturbation consists of a water jet injected by a solenoid valve, placed 100 D downstream of the pipe entrance, which is run at frequencies ranging from f = 0.2 Hz to f = 2.5 Hz. The valve opening time, frequency and water discharge are independently controlled. The duration of each jet is $t_\mathrm{jet}$ = 0.11 s $\approx$ 2.5 D/U and the amount of water injected is 0.15 ml per discharge which corresponded to less than one-hundredth of the pipe discharge at Re=2000. At the distance  of 400 D downstream the perturbation point, the number of induced puffs, $N_\mathrm{puffs}$, is counted during $300$ s  $\approx$ 7500 $D$/$U$. For each frequency of the input disturbance the experiments are repeated five times to extend the observation period to $t_\mathrm{obs}$ $\approx$ 37500 $D$/$U$. In the following the temporal spacing of perturbations is converted into an axial spacing of pulses by multiplication with the mean velocity U, assuming that perturbations are advected with the mean flow speed.
This assumption is supported by a previous study \cite[]{Lozar2} where we show that isolated puffs propagate only $\sim$ 5 $\%$ slower than the mean flow at these Reynolds numbers.

\begin{figure}
\centering
\includegraphics[width=13.5 cm]{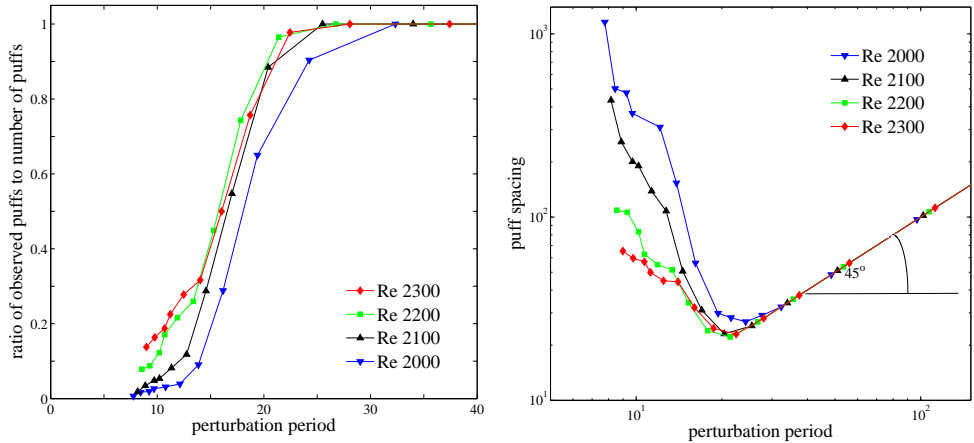}
\vspace{-0.5em}
 \caption{ $(a)$ Ratio of number of puffs to number of perturbations  vs. perturbation period.
 $(b)$Distance between puffs vs. perturbation period. }\vspace{-0.5em} 
\label{combined1}
\end{figure}

The ratio of the number of observed puffs to the number of perturbation pulses is shown in figure \ref{combined1}(a) as a function of the perturbation period. For large axial spacing (small frequencies) each perturbation results in a single puff. However once the puff spacing is below $\sim$ 20-25 D, new generated puffs interact and annihilate each other and the ratio of created puffs to perturbations drops below one.  Figure \ref{combined1}(b) shows the distance between the generated puffs in response to the perturbation period. From figure \ref{combined1}(b) it is again evident that the rate of puff generation is in synchrony with the rate of injected perturbation for large axial spacing and that the puff spacing is identical to the spacing preselected by the perturbation. Once puffs are created at too short distances ($<20$-$25 D$) the number of puffs observed downstream drops resulting in an increase of the distance between observed puffs. The minimum of each curve in figure \ref{combined1}(b) corresponds to the perturbation frequency generating the maximum number of puffs which can sustain themselves in a given length of the pipe. The corresponding distance will in the following be referred to as the optimum spacing as it provides the maximum turbulent fraction. For times between perturbations shorter than the minimum one (high frequencies), the generation of puffs decreases quite sharply for the lower Reynolds numbers. For instance at Re = 2000, only one puff each $1000\,$D is observed at high perturbation frequencies (1/f=7$D$/$U$). At these frequencies perturbed regions do not always merge to form a single puff but
instead quite frequently they annihilate each other. This decrease of generated puffs is smaller for the higher Reynolds numbers where the distance between puffs seems to saturate for large frequencies.

\begin{figure}
\centering
\includegraphics[width=7.5cm]{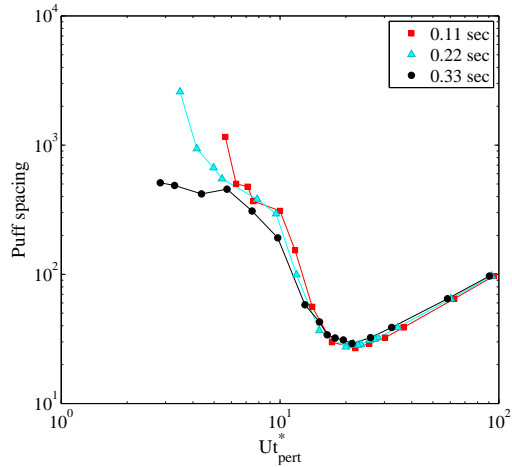}
\vspace{-0.4em}
\caption{Impact of valve opening time on puff spacings. Three different valve timings (0.11, 0.22 \& 0.33 sec) were used. In the x-axis  $t_{pert}^{*}$ is the time between pulses taking valve opening time into account (see text)}
\vspace{-0.4em} 
\label{valvetiming}
\end{figure}

 Next we investigate whether the generation of puffs depends on the intricacies of the jet injection method i.e. the water injected per perturbation and the valve opening time. The puff spacing is found to remain unchanged when the jet discharge is varied from 0.15 ml/discharge to 0.25 ml/discharge. The impact of  the jet duration on puff spacing has also been investigated. Here the valve opening time is varied from $t_{valve}$ = 0.11 s to 0.22 s and to 0.33 s for Re=2000. From figure \ref{valvetiming} it is evident that all curves collapse when plotted as a function of the time between perturbation pulses i.e. the time between the end of one jet and the beginning of the next one ($t_\mathrm{pert}^{*} = t_\mathrm{pert} $-$ t_\mathrm{valve})$. This highlights that the relevant time is not the period of the perturbation but the time for which the flow is unperturbed. Since the minimum of all curves is the same, the optimum puff spacing is independent of the details of the perturbation.

\begin{figure}
\centering
\includegraphics[width=13.5cm]{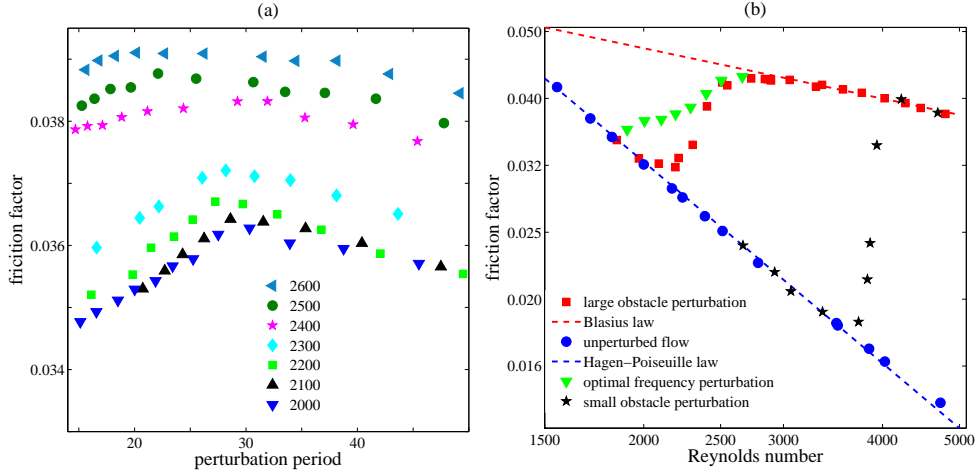}
\vspace{-0.5em}
\caption{$(a)$ Friction factor vs. perturbation period. $(b)$ Friction factor vs. Reynolds number using different perturbation types as indicated in the legend.}
\vspace{-0.5em} 
\label{combined}
\end{figure}

In the following the friction factor is measured for three cases: for a flow without perturbation, when the flow is tripped by an obstacle and for the periodic perturbation described earlier. The friction factor quantifies the friction losses and it is defined in a pipe as $ f=\Delta p / (1/2 \rho U^2 L/D)$, where $\Delta p$ is the pressure drop over a length $L$ and $\rho$ is the fluid density. When the flow is not perturbed, the friction factor closely follows the Hagen-Poiseuille law $f = 64/Re$. For a fully turbulent flow the friction factor lies on the Blasius curve $f=0.3164Re^{-0.25}$. On the other hand, in the transitional regime the friction factor strongly depends on initial conditions as well as experimental imperfections. To demonstrate this dependence, friction factors resulting from two different obstacles are shown in figure \ref{combined}(b)(red squares and grey open circles). For the small obstacle transition sets in at $\Delta$Re $\approx$ 1500 larger than for the large obstacle. Depending on the perturbation amplitude and the observation point any transitional friction factor (i.e points between the Hagen-Poiseuille and the Blasius curves) can be observed for Reynolds number Re$>$2000. 

In order to investigate the dependence of the friction factor on initial conditions we study the pressure drop for varying frequencies of the jet perturbation. The flow is again periodically perturbed as described earlier and 100 D downstream of the perturbation point the pressure drop is measured over a length of 770 D (the pipe length was increased by 5 meter for this purpose) with a Validyne DP45 pressure sensor. The pressure drop over this length lies in the optimal working range of the DP45 sensor resulting in an accuracy better than one percent for the pressure readings. Starting from small perturbation frequencies, i.e. large spacing, the friction factor is found to first increase with frequency and then reaches a maximum at the optimum puff spacing (figure \ref{combined}(a)). Beyond this point the perturbed regions start to interact resulting in a lower density / packing fraction of puffs. The maximum packing fraction data provides an upper bound for the friction factor for a given Reynolds number, shown by the green triangles in figure \ref{combined}(b). These data therefore provide a well defined connection between the Hagen-Poiseuille law and Blasius law. While the maximum packing fraction is clearly differentiated for Re$<$2300 the curves become flatter when increasing the Reynolds number until the point that the curves are almost flat and only a very weak interaction can be detected ($Re=2500$). The fading of the maximum indicates that a fully turbulent flow state is approached. This is in qualitative agreement with recently suggested transition points to fully turbulent flow at $Re=2550$ to 2600 \cite[]{Lozar, Moxey} although from our data no criticality can be inferred.

\subsection{Reduction experiments} 
In the following we would like to investigate if in analogy to TCF and PCF  "large-wavelength instability" scenario \cite[]{Prigent, Prigent1, Barkley1} can also be observed in pipe flow. In the above studies it has been suggested that when the Reynolds number is reduced from the fully turbulent regime, an instability occurs which gives rise to periodic laminar-turbulent patterns. If the same holds for pipe flow, its expected that upon reduction of Re from the fully turbulent regime, laminar gaps open up at regular intervals giving rise to a periodic laminar turbulent pattern. We therefore initially created a uniform turbulent flow at Re=4500 by putting an obstacle into the pipe directly behind the entrance and subsequently reduced the Reynolds number to the intermittent regime. By pressure measurements we verified that the friction factor lies on the Blasius curve (figure \ref{combined}b) and hence that the flow is initially turbulent throughout the pipe. Next a bypass valve was opened allowing half the total discharge to flow through a parallel pipe (not shown in figure \ref{setup}). By doing so, the Reynolds number in the pipe decreases to the desired value. Care was taken to avoid air pockets or elastic components in the setup which would cause damped oscillations in the flow rate during the reduction. Flow visualization and monitoring of the outflow angle at the pipe exit showed that the flow changes smoothly between both Reynolds numbers during approximately four seconds ($\approx$ 100 D/U). 

\begin{figure}
 \centering
 \begin{tabular}{cc}
   $(a)$ & $(b)$ \\
   \includegraphics[width=0.48\linewidth,clip=]{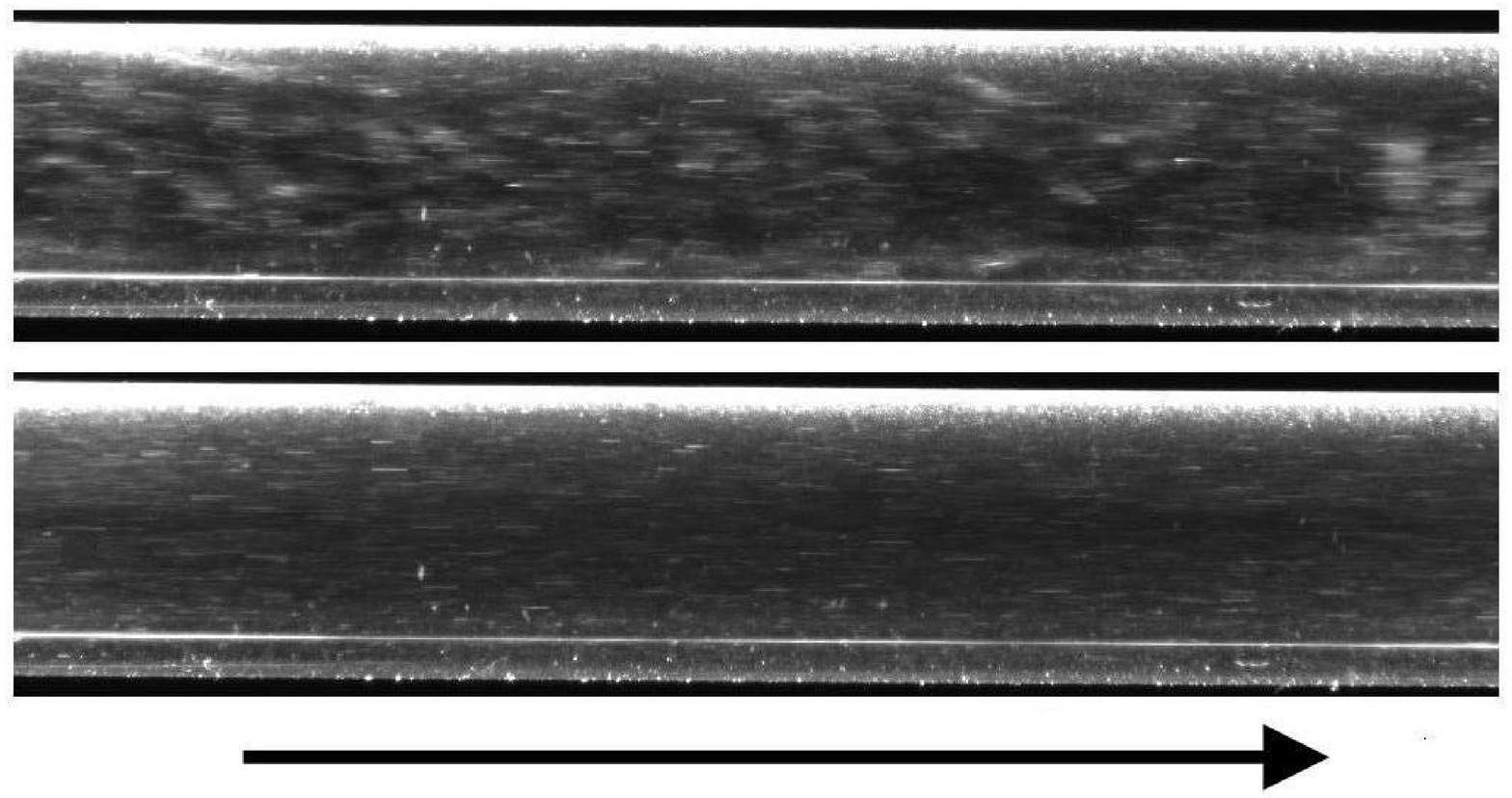}&
   \includegraphics[width=0.48\linewidth,clip=]{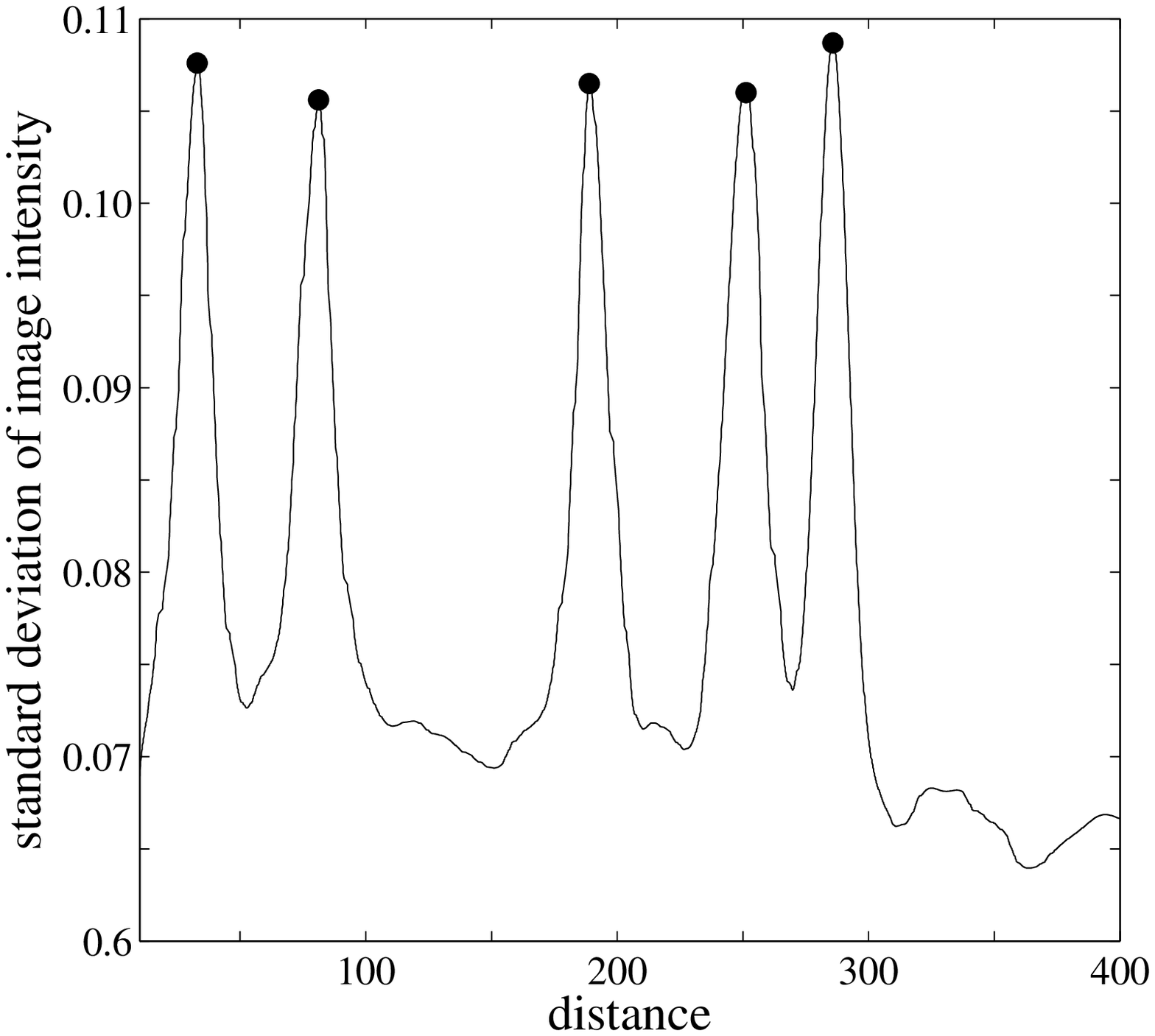}
 \end{tabular}
 \caption{ $(a)$ Flow visualization at $Re=2000$ in turbulent(upper image) and laminar flow (lower image). Images display a pipe segment of 5-6 D long. Flow direction is from left to right. One reduction experiment is shown in movie 1 available in the online version of the paper.  $(b)$ Standard deviation of image intensity vs. reconstructed pipe position at $Re=2100$. Summits of the peaks (black dots) indicate the position of a turbulent spot.}
  \label{combined2}
\end{figure}

Images were captured at 10 Hz with a CCD camera (Matrix Vision Blue Fox 121G) placed 500 D downstream of the entrance. Figure \ref{combined2}(a) shows typical snapshots for a laminar and a turbulent flow. The standard deviation in the turbulent image intensity is greater than in the laminar flow which allows for a simple identification of the turbulent regions. The laminar turbulent pattern in the pipe was reconstructed using the Taylor's hypothesis at the Reynolds numbers of the experiments. Generally it was found that after reduction the flow rapidly ($<$100 $D$/$U$) adjusts to the new Reynolds number resulting in a fixed pattern of equilibrium puffs separated by laminar flow. Consequently, initial 4 seconds (100 time units) of each signal were disregarded to allow the flow to establish the equilibrium pattern allowing a measurement time of 400 $D/U$. Once established, this pattern shows little dynamics on the time scale of the experiment for Re$<$2400. This was confirmed by a second set of measurements where the waiting time was increased to 300 D/U resulting in a remaining measurement time of 200 D/U (data not shown). Both sets of measurements (200 D/U and 400 D/U) resulted in the same size distributions (within experimental error). The dynamics of puffs only appear to be relevant on much longer time scales, puff splitting in pipes of similar length has been observed \cite[]{nishi}for Re$>$2450 and the probability of an isolated puff to decay during 500 time units at $\mathrm{Re}=2000$ is around $5\cdot 10^{-4}$ \cite[]{Hof3, Hof, Avila}. A sample signal at Re 2100 is presented in figure \ref{combined2}(b). The data has been postprocessed by smoothing the original signal.The peaks in figure \ref{combined2}(b) represent the presence of turbulent puffs and the intermediate valleys correspond to the laminar regions. The summits of the peaks were traced and represent the puff positions. After 600 time units all the fluid which initially had been in the fully turbulent state had been advected past the pipe exit and the entire flow is laminar.

\begin{figure}
\centering
\includegraphics[width=13.5cm]{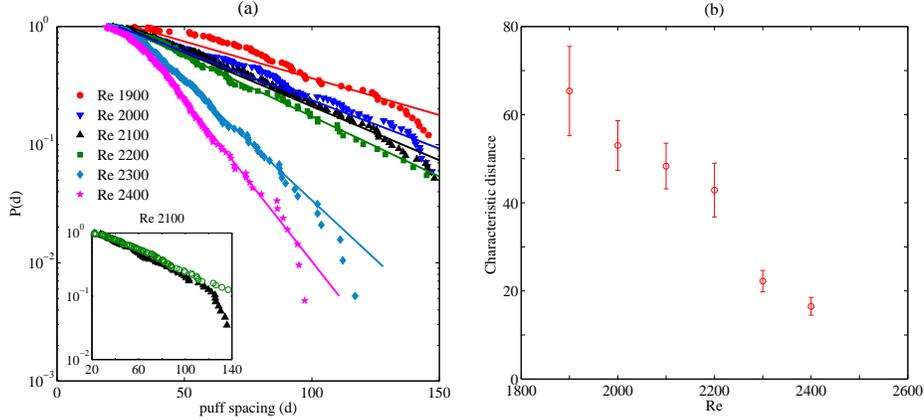}
\vspace{-0.5em}
\caption{$(a)$ Probability that puffs are separated by a distance greater than $d$ after reducing the Reynolds number from $Re = 4500$. At the initial $Re$ the flow is fully turbulent. The inset compares experiments with two initial Reynolds numbers (black triangles for $Re=4500$ and green circles for $Re=6000$).  Note that deviations for d $>$100 were due to the finite sample size.
 $(b)$ Characteristic distance $\delta$ from equation \ref{eqn} as function of the Reynolds number.Error bars correspond to 95 $\%$ confidence interval.}
\vspace{-0.5em} 
\label{combined3}
\end{figure}

For each Re 30 to 50 experiments were performed and this resulted in a sample size of about 200 puff spacings. In general, the puff spacing decreased with increasing Reynolds number. Typically 4-5 puffs appeared for $\mathrm{Re}=2000$ whereas the number of puffs for $\mathrm{Re}=2400$ was 8-10 for the same time interval. From the data we determined  the cumulative probability $P(d)$ that two adjacent puffs are separated (peak to peak distance) by a distance d or larger. For a spatially periodic pattern such cumulative distribution should show a sharp drop at the wavelength of the pattern. Instead, as shown in figure \ref{combined3}(a), the data falls on exponential tails. The fact that the spacing between the puffs is highly irregular is also seen in the example shown in figure \ref{combined2}(b). The exponential distributions ( figure \ref{combined3}(a)) do not cross zero but reach $P(d)$=1 at some value $d$=$d_{min}$ $>$ 0 where $d_{min}$ is the minimum puff spacing. The distribution takes the form of :

\begin{equation}
\label{eqn}
P(d) \propto  e^{\frac{-d}{\delta}} \mbox{ for } d > d_{0} 
\end{equation}

where $d_{0}$ is related to the distance from which the exponential distribution applies and $\delta$ is the characteristic scale of the exponential distribution. Note that $\delta$$+$$d_{0}$ is the mean distance between the puffs. The statistical analysis of an exponential distribution which only applies from a distance $d_{0}$ has been performed using techniques described in \cite{Avila}. This probability function implies that the spacing between two puffs is always larger than $d_{0}$ but otherwise arbitrary and that it is not influenced by the spacing between puffs further up and downstream. Therefore the existence of a preferred spacing or wavelength is ruled out. 

For the lower Reynolds numbers ($\mathrm{Re}=1900$-2200) the exponential distribution applies from the minimum puff distance ($d_0 = d_{min}$) showing that the appearance of a puff is not influenced by any upstream or downstream puff. Note that for $Re = 2300$ and 2400 there is a slight initial rounding indicating that size distributions only become random for puff spacings slightly larger than $d_{min}$. The deviations from the exponential distributions at long distances on the other hand can be attributed to the finite size of the pipe.

To rule out any effects of initial Reynolds number on the reduction experiments, a second set of experiments was carried out where the Reynolds number was reduced from a turbulent flow at Re=6000. Excellent agreement is found with the reduction experiments from Re =4500 (see the inset in figure \ref{combined3}(a)). Note that the deviations at the end of the tails  carry little statistical weight and can be attributed to finite sample size. This indicates that the size distribution of the observed flow pattern does not depend on the initial intensity of the turbulence but only on the Reynolds number established after reduction. 

The characteristic distance $\delta$, from equation \ref{eqn}, clearly decrease with Re (figure \ref{combined3}(b)). As expected turbulent spots are more closely spaced with the increase of the Reynolds number. While the quality of the data does not allow to determine the exact functional dependence the characteristic distances appear to approach zero for Re $>\approx$ 2500, where maximum puff packing should be achieved. This point is very close to the transition to uniform turbulence ($Re_c = 2550-2600$) and to the Reynolds number where the puff interaction diminishes (figure \ref{combined}(a)).

\section{Discussion \& Conclusions}

\begin{figure}
\centering
\includegraphics[width=7.5cm]{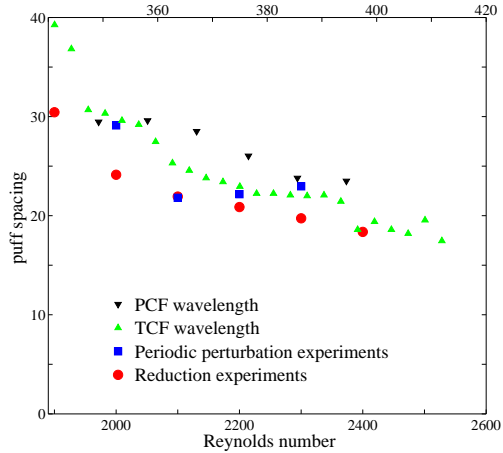}
\vspace{-1em}
\caption{Comparison of minimum puff spacing (Reduction experiments) and optimum puff spacing (Periodic perturbation experiments) with laminar turbulent wavelengths in transition regimes of Taylor-Couette and Plane-Couette flow from \cite{Prigent}. The upper x axis represents the Reynolds
 number used for Taylor Couette flow and Plane Couette flow and the lower x axis represents the Reynolds number used in our experiments.}
\vspace{-1em} 
\label{comparison}
\end{figure}

The experiments presented in this paper establish the existence of a minimum distance between turbulent patches in pipe flow. In the first set of experiments an optimum distance corresponding to the densest packing fraction of puffs has been determined with the aid of a periodic perturbation.  Specifically we measured the maximum achievable friction factor in the transition regime as a function of Re. Typically the friction factor in this regime strongly depends on initial conditions and varies from one experiment to another. Our data gives an upper bound for the sustainable friction factor  and provides a well defined link between the Hagen-Poiseuille and the Blasius law. In the second set of experiments, where no periodicity was imposed externally, a minimum spacing has been observed.  In Figure \ref{comparison} we compare the minimum distance from the reduction experiments with the optimum distance from the periodic perturbation experiments as function of the Reynolds number. Both data sets overlap indicating that the minimum spacing is a property intrinsic to the flow, independent of initial conditions or flow perturbations. The existence of a minimum spacing in pipe flow can be explained following the arguments presented in \cite{Hof1} where it has been pointed out that the instability mechanism sustaining a puff relies on an inflection point at the rear of the puff. When an upstream puff appears the inflection point behind the original puff is distorted. Eventually the downstream puff decays if the distortion is strong enough. The nature of the instability at the rear end of the puff has also been discussed by \cite{Kida} and \cite{Duguet}. Similar to the arguments given in an earlier study by \cite{Bandyopadhyay} they emphasize the relevance of Kelvin Helmholtz vortices created at the rear interface to the puff sustenance. 

It should be pointed out that the active region of a turbulent puff is approximately 10 D long \cite[]{Hof1}. Further downstream the fluctuations have disappeared and the profile gradually recovers the parabolic shape until the preceding puff is encountered. This aspect is similar to the observations in stripe patterns in PCF \cite[]{Barkley2}. The turbulent eddies are localised in the turbulent stripes which approximately have a width of 10 wall spacings. Between two stripes the flow relaminarizes but does not quite reach the fully developed laminar profile. A comparison of the minimum puff spacing in pipes to the wavelength measured in PCF and TCF laminar turbulent patterns \cite[]{Prigent} surprisingly shows that the wavelength closely matches the interaction distance found in pipes when the Reynolds number is rescaled so that the intermittent region for the different flows overlap. The wavelength have been calculated from the spanwise and streamwise spacings measured by \cite{Prigent1}. Banded laminar-turbulent patterns with similar wavelengths have been reported in TCF, PCF, torsional Couette flow and plane Poiseiulle flow \cite[]{Barkley2}. Our measurements extend this similarity to the case of pipe flow where the typical length scale does not appear as a wavelength but as a minimum or optimum distance.We hope that this observation will stimulate further studies to clarify similarities and differences between these flows. 

Our results indicate that in pipes the spacing between adjacent turbulent regions is a consequence of nearest neighbour interaction process and that no long range instability as suggested for PCF and TCF \cite[]{Prigent,Prigent1} is required. It should also be pointed out that as a recent study shows \cite[]{Kerstin} the spreading and splitting of puffs takes place on time scales much larger than those available in typical pipe experiments. It is therefore possible that in the asymptotic time limit a more regular puff spacing is reached, which may be closer to the maximum packing fraction. If so this flow pattern would result from nearest neighbour interaction and not manifest a least stable wavelength of an instability.

\textbf{Acknowledgments:} We thank Dr. Marc Avila and the reviewers for their valuable suggestions in improving the quality of the paper.

\bibliography{puff2}

\begin{thebibliography}{10}

\bibitem{Kerswell}
R.~Kerswell, ``Recent progress in understanding the transition to turbulence in
  a pipe,'' {\em Nonlinearity}, vol.~18, pp.~17--44, 2005.

\bibitem{Eckhardt2}
B.~Eckhardt, T.~Schneider, B.~Hof, and J.~Westerweel, ``Turbulence transition
  in pipe flow,'' {\em Annual Review of Fluid Mechanics}, vol.~39,
  pp.~447--468, 2007.

\bibitem{Hof4}
B.~Hof, C.~Vandoorne, J.~Westerweel, F.~Nieuwstadt, H.~Faisst, B.~Eckhardt,
  H.~Wedin, R.~Kerswell, and F.~Waleffe, ``Experimental observation of
  nonlinear travelling waves in turbulent pipe flow,'' {\em Science}, vol.~305,
  pp.~1594--1598, 2004.

\bibitem{Reynolds}
O.~Reynolds, ``An experimental investigation of the circumstances which
  determine whether the motion of water shall be direct or sinuous, and of the
  law of resistance in parallel channels,'' {\em Proc. R. Soc. London A},
  vol.~35, pp.~84--99, 1883.

\bibitem{Rotta}
J.~Rotta, ``Experimenteller beitrag zur enstehung turbulent stroemung im
  rohr,'' {\em Ing-Arch}, vol.~24, pp.~258--251, 1956.

\bibitem{Wygnanski1}
I.~Wygnanski and F.~Champagne, ``On transition in a pipe. part 1. the origin of
  puffs and slugs and the flow in a turbulent slug,'' {\em J. Fluid Mech.},
  vol.~59, pp.~281--335, 1973.

\bibitem{Wygnanski2}
I.~Wygnanski, M.~Sokolov, and D.~Friedman, ``On transition in a pipe. part 2.
  the equilibrium puff,'' {\em J. Fluid Mech.}, vol.~69, pp.~283--304, 1975.

\bibitem{Schneider}
T.~Schneider, J.~Gibson, and J.~Burke, ``Snakes and ladders: Localized
  solutions of plane couette flow,'' {\em Phys. Rev. Lett.}, vol.~104,
  p.~104501, 2010.

\bibitem{Mellibovsky}
F.~Mellibovsky, A.~Meseguer, T.~Schneider, and B.~Eckhardt, ``Transition in
  localized pipe flow turbulence,'' {\em Phys. Rev. Lett.}, vol.~103,
  p.~054502, 2009.

\bibitem{Willis}
A.~Willis and R.~Kerswell, ``Turbulent dynamics of pipe flow captured in a
  reduced model: puff relaminarization and localized 'edge' states,'' {\em J.
  Fluid Mech.}, vol.~619, pp.~213--233, 2009.

\bibitem{Duguet}
Y.~Duguet, A.~P. Willis, and R.~R. Kerswell, ``Slug genesis in cylindrical pipe
  flow,'' {\em J. Fluid Mech.}, vol.~663, pp.~180--208, 2010.

\bibitem{Coles}
D.~Coles, ``Transition in circular couette flow,'' {\em J. Fluid Mech.},
  vol.~21, pp.~385--425, 1965.

\bibitem{Prigent}
A.~Prigent, G.~Gregoire, H.~Chate, O.~Dauchot, and W.~Sarloos, ``Large-scale
  finite-wavelength modulation within turbulent shear flows,'' {\em Phys. Rev.
  Lett.}, vol.~89, p.~014501, 2002.

\bibitem{Prigent1}
A.~Prigent, G.~Gregoire, H.~Chate, and O.~Dauchot., ``Long-wavelength
  modulation of turbulent shear flows,'' {\em Physica D}, vol.~174,
  pp.~100--113, 2003.

\bibitem{Barkley1}
D.~Barkley and L.~Tuckerman, ``Computational study of turbulent laminar
  patterns in couette flow,'' {\em Phys. Rev. Lett.}, vol.~94, p.~014502, 2005.

\bibitem{Cros}
A.~Cros and P.~L. Gal, ``Spatiotemporal intermittency in the torsional couette
  flow between a rotating and a stationary disk,'' {\em Phys. Fluids}, vol.~14,
  pp.~3755--3765, 2002.

\bibitem{Savas}
O.~Savas, ``On flow visualization using reflective flakes,'' {\em J. Fluid
  Mech.}, vol.~152, pp.~235--248, 1985.

\bibitem{Gauthier}
G.~Gauthier, P.~Gondret, and M.~Rabaud, ``Motion of anisotropic
  particles:application to visualization of three-dimensional flows,'' {\em
  Phys. Fluids}, vol.~10, pp.~2147--2154, 1998.

\bibitem{Hof1}
B.~Hof, A.~de~Lozar, M.~Avila, X.~Tu, and T.~M. Schneider, ``Eliminating
  turbulence in spatially intermittent flows,'' {\em Science}, vol.~327,
  pp.~1491--1494, 2010.

\bibitem{Lozar2}
A.~de~Lozar and B.~Hof, ``An experimental study of the decay of turbulent puffs
  in pipe flow,'' {\em Phil. Trans. of Royal Soc. A}, vol.~367, pp.~589--599,
  2009.

\bibitem{Lozar}
A.~de~Lozar and B.~Hof, ``Universality at the onset of turbulence in shear
  flows,'' {\em arXiv:1001.2481}, 2010.

\bibitem{Moxey}
D.~Moxey and D.~Barkley, ``Distinct large-scale turbulent-laminar states in
  transitional pipe flow,'' {\em PNAS}, vol.~107, pp.~8091--8096, 2010.

\bibitem{nishi}
M.~Nishi, B.~{\"U}nsal, , F.~Durst, and G.~Biswas, ``Laminar-to-turbulent
  transition of pipe flows through puffs and slugs,'' {\em J. Fluid Mech.},
  vol.~614, pp.~425--446, 2008.

\bibitem{Hof3}
B.~Hof, J.~Westerweel, T.~M. Schneider, and B.~Eckhardt, ``Finite lifetime of
  turbulence in shear flows,'' {\em Nature}, vol.~443, pp.~59--62, 2006.

\bibitem{Hof}
B.~Hof, A.~de~Lozar, D.~J. Kuik, and J.~Westerweel, ``Repeller or attractor?
  selecting the dynamical model for the onset of turbulence in pipe flow,''
  {\em Phys. Rev. Lett.}, vol.~101, p.~214501, 2008.

\bibitem{Avila}
M.~Avila, A.~P. Willis, and B.~Hof, ``On the transient nature of localized pipe
  flow turbulence,'' {\em J. Fluid Mech}, vol.~646, pp.~127--136, 2010.

\bibitem{Kida}
M.~Shimizu and S.~Kida, ``A driving mechanism of a turbulent puff in pipe
  flow,'' {\em Fluid Dyn. Res.}, vol.~41, pp.~1--27, 2009.

\bibitem{Bandyopadhyay}
P.~Bandyopadhyay, ``Aspects of the equilibrium puff in transitional pipe
  flow,'' {\em J. Fluid Mech.}, vol.~163, pp.~439--458, 1986.

\bibitem{Barkley2}
D.~Barkley and L.~Tuckerman, ``Mean flow of turbulent$-$ laminar patterns in
  plane couette flow,'' {\em J. Fluid Mech.}, vol.~576, pp.~109--137, 2007.

\bibitem{Kerstin}
K.~Avila, D.~Moxey, A.~.de Lozar, M.~Avila, D.~Barkley, and B.~Hof, ``Onset of
  sustained turbulence in pipe flow,'' {\em Under submission}, 2011.

\end{thebibliography}
\bibliographystyle{ieeetr}
\end{document}